\documentclass[prd,nofootinbib,tightenlines,floatfix,showpacs,amssymb,aps,superscriptaddress]{revtex4-1}
\usepackage{graphicx}
\usepackage{dcolumn}
\usepackage{bm}
\usepackage{epsfig}
\usepackage{amsmath}
\usepackage{xcolor}
\usepackage{threeparttable}
\usepackage{amsfonts}
\usepackage{amssymb}
\usepackage{amsmath}
\usepackage{graphicx}
\usepackage{color}
\usepackage[colorlinks,linkcolor=blue,anchorcolor=green,citecolor=blue]{hyperref}
\usepackage{latexsym}
\usepackage{longtable}
\usepackage{changepage}
\usepackage{tabulary}

\def\aap{A\&A}

\def\apj{ApJ}
\def\apjl{ApJL}
\def\apjs{ApJS}

\def\apss{Ap\&SS}

\def\caa{Chinese Astronomy and Astrophysics}

\def\nat{Natur}
\def\natA{Nature Astronomy}

\def\mnras{MNRAS}
\def\mpla{Modern Physics Letters A}
\def\pasa{Publications of the Astronomical Society of Australia}

\def\pasj{PASJ}

\def\prd{Phys.~Rev.~D}

\def\prl{Phys.~Rev.~Lett.}
\def\PrPNP{Progress in Particle and Nuclear Physics}
\def\raa{Research in Astronomy and Astrophysics}

\def\Sci{Science}

\def\ArX{ArXiv e-prints}

\begin{document}

\title{Statistical properties of the repeating fast radio burst source FRB 121102}

\author{Bing Li}
\email{libing@ihep.ac.cn}
\affiliation{School of Astronomy and Space Science, Nanjing University, Nanjing 210023, China}
\affiliation{Laboratory for Particle Astrophysics, Institute of High Energy Physics, Beijing 100049, China}
\affiliation{Key Laboratory of Particle Astrophysics, Chinese Academy of Sciences, Beijing 100049, China}

\author{Long-Biao Li}
\email{lilongbiao0919@163.com}
\affiliation{School of Mathematics and Physics, Hebei University of Engineering, Handan 056005, China}

\author{Zhi-Bin Zhang}
\email{z-b-zhang@163.com}
\affiliation{College of Physics and Engineering, Qufu Normal University, Qufu 273165, China}

\author{Jin-Jun Geng}
\email{gengjinjun@nju.edu.cn}
\affiliation{School of Astronomy and Space Science, Nanjing University, Nanjing 210093, China}
\affiliation{Key Laboratory of Modern Astronomy and Astrophysics, Ministry of Education, Nanjing 210023,China}

\author{Li-Ming Song}
\email{songlm@ihep.edu.cn}
\affiliation{Laboratory for Particle Astrophysics, Institute of High Energy Physics, Beijing 100049, China}
\affiliation{Key Laboratory of Particle Astrophysics, Chinese Academy of Sciences, Beijing 100049, China}

\author{Yong-Feng Huang}
\email{hyf@nju.edu.cn}
\affiliation{School of Astronomy and Space Science, Nanjing University, Nanjing 210093, China}
\affiliation{Key Laboratory of Modern Astronomy and Astrophysics, Ministry of Education, Nanjing 210023,China}

\author{Yuan-Pei Yang}
\affiliation{Kavli Institute for Astronomy and Astrophysics, Peking University, Beijing 100871, China}
\affiliation{National Astronomical Observatories, Chinese Academy of Sciences, Beijing 100012, China}

\date{\today}

\begin{abstract}
  Currently, FRB 121102 is the only fast radio burst source that was observed to give out
  bursts repeatedly. It shows a high repeating rate, with more than one hundred bursts being spotted, but
  with no obvious periodicity in the activities. Thanks to its repetition, the source was well
  localized with a subarcsecond accuracy, leading to a redshift measurement of about 0.2.
  FRB 121102 is a unique source that can help us understand the enigmatic nature of fast radio bursts.
  In this study, we analyze the characteristics of the waiting times between bursts from FRB 121102.
  It is found that there is a clear bimodal distribution for the waiting times.
  While most waiting times cluster at several hundred seconds, a small portion of
  the waiting times are strikingly in the range of 2 --- 40 millisecond.
  More interestingly, it is found that the waiting time does not correlate
  with the burst intensity, either for the preceding burst or for the subsequent burst.
  It strongly indicates that the repeating bursts should be generated by some external
  mechanisms, but not internal mechanisms. As a result, the models involving collisions
  between small bodies and neutron stars could be competitive mechanisms for this interesting source.
\end{abstract}

\maketitle

\section{Introduction}

Fast radio bursts (FRBs) are intense extragalactic radio bursts with a flux of $Jy$ level and
with short durations of about several millisecond. Their origin and trigger mechanism is still
an open question.
The first FRB event (010724) was once named according to the discoverer as ``Lorimer Burst''
\citep{Lorimer2007Sci}. After about one decade, more than 60 FRBs have been found by
terrestrial radio telescopes \citep{Petroff2016PASA}\footnote{Petroff et al.2016 catalogued and
updated all published sample of FRBs, see their online catalog--http://www.frbcat.org}.
FRBs have anomalously high dispersion measures (DMs), significantly exceeding the
expected Milky Way contribution along the line of sight, which is contrary to Galactic pulsars
\citep{Thornton13Sci,Champion16MN,Cordes16arX,Bannister17ApJ,YaoJM17ApJ,Caleb18MN478,
Shannon18Nat,Ravi19MN} (but note that low DM has also been hinted
from two FRBs \citep{Ravi16Sci,Petroff19MN} and a candidate (FRB 141113) \citep{Patel18arX}).
It indicates that FRBs are of extragalactic or even cosmological origin, rather than of
Galactic origin \citep{Dolag15MN,Caleb16MN458,Katz16ApJ818,Yang17ApJ839,WeiJJ18ApJ860}.
Their bright millisecond radio pulses with a high scattering effect point to extremely high
brightness temperatures, which implies that luminous coherent emission processes
around compact objects should be involved, which might include the curvature radiation
(also known as the ``antenna'' mechanism) \citep{Ghisellini18AA613,Katz18MN481,YangYP18ApJ868},
or the synchrotron maser emission \citep{Lyubarsky14MN,Ghisellini17MN,Waxman17ApJ,LongK18arX}.
However, the exact mechanism is still enigmatic and greatly debated \citep{Katz16MPLA,Katz18PrPNP,Lu18MNR477}.

Numerous theoretical models have been proposed to account for such
an enigmatic class of radio transients. The number of models is almost comparable to that of the observed
FRBs up to date \citep{Caleb18NatA,PenUL18Nat}.
Most of those progenitor models involve compact objects (e.g., Neutron Stars (NS),
Black Holes (BH) and White Dwarfs(WD)) outside of the Milky Way. Popular mechanisms include:
mergers of binary NS \citep{Totani13PASJ,WangJS16ApJ,Dokuchaev17},
or binary WD \citep{Kashiyama2013ApJ}, or NS-WD binary \citep{GuWM16ApJ,LiuX18APSS},
or NS/WD-BH binary \citep{Mingarelli15ApJ,Abramowicz18ApJ,LiLB18RAA}, the interactions of
pulsar-BH systems \citep{Bhattacharyya17arX} or Kerr-Newman BH-BH \citep{ZhangB16ApJ827,LiuT16ApJ},
collapse of compact objects (e.g., collapse of NS \citep{Falcke14AA,ZhangB14ApJ,Fuller15MN,
Punsly16MN,Shand16RAA}, collapse of strange star crust \citep{ZhangY18ApJ}), giant pulses/flares
from magnetars or young pulsars \citep{Keane12MN,Cordes16MN,Connor16MN,Lieu17ApJ}, and other
supernovae interrelated theories \citep{Popov2007arX,Popov2013arX,Lyubarsky14MN,Murase16MN,Beloborodov17}).
A few other models include: mechanisms related to active galactic nuclei(AGN) and
Kerr-BH or Strange Star interactions \citep{Romero16PRD,Vieyro17AA,Gupta17arX,Katz471MN92,ZhangY18},
collisions between NSs and small bodies \citep{MottezZar2014AA,GengHuang2015ApJ,HuangYF16ASPC,Dai2016ApJ829},
Axion star collides with a NS/BH \citep{Iwazaki15PRD,Raby16PRD,Iwazaki17arX}, interactions
between Axions and compact bodies \citep{Tkachev15,Rosa18PRL,vanWaerbeke18arX},
explosions such as starquakes \citep{WangWY18ApJ},
primordial BHs/Planck stars collapse to form white holes \citep{Rarrau14PRD,Rarrau18PRD},
lightning/wandering in pulsars \citep{Katz17MN469,Katz16ApJ818},
or NS combing \citep{ZhangB17ApJ836,ZhangB18ApJ854}, and so on.
In short, while the true mechanisms of FRBs are still unclear, it is possible that multiple
populations of FRBs might exist \citep{Rane17JApA,Michilli18Nat,Palaniswamy18ApJ}.
For a summary of FRB progenitor models, see the recent article of Platts et al. \citep{Platts18arX}
and the online version of a tabulated summary\footnote{http://frbtheorycat.org}.

In general, the above models could be grouped into two different categories \citep{Dai17ApJ838,LiLB18Caa},
i.e. catastrophic models and non-catastrophic models. Following this idea, a key issue is then
whether the observed FRBs are repetitive or not. Strangely enough, among the more than 60 FRB
sources, only one event, i.e. FRB 121102 is observed to repeatedly burst out, which actually
has produced hundreds of bursts so far.
For all the other FRB sources, no indication of repetition was observed although
long-term monitoring had been extensively carried
out \citep{Lorimer2007Sci,Thornton13Sci,Petroff15MN,Ravi15ApJ,
Ravi16Sci,Petroff17MN,Bhandari18MN,Shannon18Nat}.
However, strictly speaking, it still remains obscure how many FRBs are repetitive
and whether their repetition is similar to the unique FRB 121102 or not. In this study,
we will mainly concentrate on the repeating event of FRB 121102.

FRB 121102 was initially discovered through the 305-m Arecibo telescope Pulsar Survey Project
(ALFA) \citep{Spitler14ApJ}. It is the first and the only known burst that has been successfully
identified to be associated with a host galaxy, a small low-metallicity star-forming dwarf galaxy at
$z = 0.19273 \pm 0.00008$ \citep{Chatterjee17Nat,Bassa17ApJ,Marcote17ApJL,Tendulkar17ApJ}.
It is also the only burst with a repeating behavior \citep{Spitler16Nat}.
Hundreds of additional bursts have been observed from the same direction toward FRB 121102 (the
position is known to sub-arcsecond precision) and the measured DMs of all these bursts (between 553 -- 569 pc $\rm cm^{-3}$)
are consistent with the first burst of 121102. These consequent observations were made discontinuously by
many terrestrial radio telescopes such as the Arecibo telescope (AO), the Green Bank Hydrogen telescope(GBT),
the Karl G. Jansky Very Large Array(VLA), the 100-m Effelsberg telescope (Eff), and the
Apertif Radio Transient System (ARTS), at multiple radio bands \citep{Spitler16Nat,Scholz16ApJ,
Chatterjee17Nat,Hardy17MN,Law17ApJ,Oostrum17ATel,Scholz17ApJ,Gajjar18ApJ,MAGIC18MN,Michilli18Nat,Spitler18ApJ,ZhangYG18ApJ}.
They all reveal no additional waveband sporadic emission \citep{Scholz16ApJ,Chatterjee17Nat,Ofek17ApJ,ZhangBB17ApJ},
and there is no evidence for any periodicity \citep{ZhangYG18ApJ}.
It is interesting that a cosmological origin is confirmed for FRB 121102 confirms with the redshift measurement.
The catastrophic models are also ruled out for this source.
FRB 121102 is associated with a variable radio source with a continuum non-thermal spectrum. It seems to be a
low-luminosity AGN or other kind of unknown peculiar source \citep{Chatterjee17Nat}. An AGN scenario thus
remains possible \citep{Romero16PRD,Vieyro17AA,Gupta17arX,Katz471MN92}, but it could also be due to other
mechanisms such as a young neutron star with pulsar wind nebula \citep{Keane12MN,Cordes16MN,Connor16MN,Lieu17ApJ}, or
a neutron star interacting with small bodies \citep{GengHuang2015ApJ,HuangYF16ASPC,Dai2016ApJ829}.

For a sporadically repeating astrophysical phenomena, the temporal information/quantities (e.g., duration,
elapsed time, periodicity, episodic time) of outbursts play an important role in understanding the
central engine, and revealing the radiation mechanism and energy dissipation processes.
For instance, the previous researches on the non-stationary Poisson process of soft gamma-ray repeaters
\citep{Cheng1996Nat,G1999ApJ,G2000ApJ}, pulsar glitches \citep{Melatos08ApJ,Haskell2016MN,Onuchukwu2016ApSS},
anomalous X-ray pulsars (AXPs) \citep{Gavriil2004ApJ,Savchenko2010AA}, X-ray flares in $\gamma$ ray burst
afterglows \citep{WangDai13Nat,Guidorzi15ApJ}, can give us helpful references.
A time-variable system always shows some kind of irregularity. Especially the time interval
between two adjacent bursts, which is known as the waiting time (here after WT or $\Delta t$),
is an important parameter that could provide valuable information on the central engine.

Previously, Gu et al.(2016) established a relation between the waiting time and
the mass transfer rate of accretion in a NS-WD binary merger \citep{GuWM16ApJ},
which can be be somewhat related to the observed behaviors of FRB 121102.
Katz(2018) found the distribution of the waiting times between bursts of FRB 121102 are apparently far
from a Poissonian form, and proposed that such a distribution may be consistent with a precessing jet
launched from a NS or a BH accretion disc \citep{Katz18MN476}.
Wang. et al.(2018) found that the energy and waiting time distributions of FRB121102 derived from
the paper \citep{Gajjar18ApJ} show an earthquake-like behavior. They argued that
those bursts could be powered by some starquake-like mechanisms \citep{WangWY18ApJ}.
Palaniswamy et al.(2018) compared 40 bursts of FRB 121102 with other non-repeating FRBs on the observed
waiting time-flux ratio plane. They found that their distribution is well separated from other FRBs
in the plane, suggesting that there could be multiple populations of FRBs \citep{Palaniswamy18ApJ}.
In this study, with more observed bursts (on the level of hundred) from the repeating source FRB 121102,
we re-analyze the waiting time statistics. Especially, we study the correlation between the waiting time
and the flux and other parameters of the repeating burster. It is expected that these new studies will help us
understand the nature of this unique source.

The structure of our paper is organized as follows.
In Section II, we collect the detailed observational data for the repeated bursts from FRB 121102.
We then calculate the waiting times for the available events.
In Section III,we derive the waiting time distribution and analysis its correlation with other parameters.
The implications of our study for the nature of the repeater is presented.
Finally, Section IV presents our conclusions and some further discussion.

\section{The observations and samples of FRB 121102}

FRB 121102 is currently the only source to exhibit repeating bursts among the FRB population.
From the literature, we have collected all the observational data of the observed bursts from this
source. An overview of the available observations is listed in Table~\ref{tab1}.
In this table, we present the date of observations, the telescope names, the starting time of the
observation, the duration of the corresponding monitoring, number of bursts observed during the
observational campaign, the number of available waiting times during the period, and references.
Note that in each observational campaign, the source usually could be monitored continuously for several
thousands (or up to several hours in several cases). A meaningful waiting time could be derived only
for two bursts that are observed in the same observational campaign. For example, if 5 bursts
were observed in a continuous observational campaign, then 4 waiting time could be derived from these
5 bursts.

Up to now, there are thirty-five continuous observation campaigns on FRB 121102 that resulted in successful
detection of at least one fast radio burst from the source. A lot of bursts are detected in Observation Campaigns 33
and 34, i.e. 46 and 47 bursts were observed, respectively. In summary, more than one burst was detected in 17
observations (see the sixth column in Table~1), providing us with 171 bursts in total.

We have calculated the waiting times for those successive bursts detected in a continuous observation.
The waiting time (WT, $\Delta t$) is defined as the time interval between two neighbouring and non-piled-up bursts
in a continuous and uninterrupted observation campaign. For the repetitive fast radio bursts from FRB 121102,
the observed profiles are composed of only a single pulse. In our calculations, we ignored the time dilation factor
of $1+z$, which is not large and will not affect our analysis. As a result, the number of waiting times
available is 136. In order to further study FRB 121102，We also collected some other important parameters 
such as the peak flux density, the fluence, the duration, etc.

\begin{table*}
  \centering
  \caption{Log of the observations on FRB 121102}
   \label{tab1}
  \begin{tabular}{lcllrcclll}
  \hline
   Obs.  &    Date      & Telescope       &  Start Time   & Obs. Length  & Number of &  Waiting time &  Reference                        \\
  number & yyyy/mm/dd   & /Receiver       & £¨hh/mm/ss£©  & $t_{obs}$(s) &  Bursts   &    number     &                                   \\   \hline
    1	   & 2012/11/02   &  AO/ALFA	      &	06:38:13      & $\sim$200    &    1	     &    0	         &  \citep{Spitler14ApJ,Spitler16Nat,Scholz16ApJ}\\
    2	   & 2015/05/17   &  AO/ALFA	      &	17:45:38	    & 1002	       &    2	     &    1	         &  \citep{Spitler16Nat}             \\	
    3 	 & 2015/06/02   &  AO/ALFA	      &	16:38:47	    & 1002	       &    2	     &    1	         &  \citep{Spitler16Nat}             \\	
    4	   & 2015/06/02   &  AO/ALFA	      &	17:48:52	    & 1002         &    6	     &    5	         &  \citep{Spitler16Nat}             \\	
    5	   & 2015/11/13   &  GBT/S-band	    &	07:42:09	    & 3000	       &    1	     &	  0	         &  \citep{Scholz16ApJ}              \\	
    6	   & 2015/11/19   &  GBT/S-band	    &	10:14:57	    & 3000	       &	  4	     &	  3	         &  \citep{Scholz16ApJ}              \\	
    7	   & 2015/12/08   &  AO/L-wide 	    &	04:43:24	    &	3625	       &	  1	     &	  0	         &  \citep{Scholz16ApJ}              \\	
    8	   & 2016/08/20   &  Eff/5 GHz	    &	09:23:40.4	  &	1200	       &	  3	     &	  2	         &  \citep{Spitler18ApJ}             \\
    9	   & 2016/08/23   &  VLA/3 GHz	    &	17:26:28	    &	3240	       &	  1	     &	  0	         &  \citep{Chatterjee17Nat,Law17ApJ} \\	
    10	 & 2016/09/02	  &  VLA/3 GHz	    &	15:52:17	    &	3240	       &	  2	     &	  1	         &  \citep{Chatterjee17Nat,Law17ApJ} \\	
    11	 & 2016/09/07	  &  VLA/3 GHz	    &	10:14:50	    &	7200	       &	  1	     &	  0	         &  \citep{Chatterjee17Nat,Law17ApJ} \\	
    12   & 2016/09/12	  &  VLA/3 GHz	    &	09:15:19	    &	7200	       &	  1	     &	  0	         &  \citep{Chatterjee17Nat,Law17ApJ} \\	
    13 	 & 2016/09/14	  &VLA/3 GHz$^{a}$  &	09:20:23	    &	7200	       &	  1      &	  0	         &  \citep{Chatterjee17Nat,Law17ApJ} \\	
    14	 & 2016/09/15	  &  VLA/3 GHz	    &	09:16:29	    &	7200	       &	  1	     &	  0	         &  \citep{Chatterjee17Nat,Law17ApJ} \\	
    15	 & 2016/09/16	  &  GBT/S-band	    &	03:59:12	    &	14452	       &	  2      &	  1	         &  \citep{Scholz17ApJ}              \\	
    16	 & 2016/09/17	  &VLA/3 GHz$^{b}$  &	08:59:20      &	7200	       &	  1      &	  0	         &  \citep{Chatterjee17Nat,Law17ApJ} \\	
    17	 & 2016/09/18	  &  GBT/S-band 	  &	04:02:15      &	14269	       &	  2	     &	  1	         &  \citep{Scholz17ApJ}              \\	
    18	 & 2016/09/18	  &VLA/3 GHz$^{a}$  &	08:59:27      &	7200	       &	  1      &	  0	         &  \citep{Chatterjee17Nat,Law17ApJ} \\	
    19	 & 2016/12/25	  &  AO4.1-4.9 GHz  &	02:48:51.998  &	6703	       &	  10	   &	  9	         &  \citep{Michilli18Nat}	           \\	
    20   & 2016/12/26	  &  AO4.1-4.9 GHz  &	02:44:21.998  &	6806	       &	  5	     &	  4	         &  \citep{Michilli18Nat}	           \\	
    21	 & 2017/01/11	  &GBT/S-band$^{c}$ &	23:13:56      &	19869	       &	  6      &	  5          &  \citep{Scholz17ApJ}              \\	
    22	 & 2017/01/12	  &AO/L-wide$^{c}$  &	01:46:27      &	6270	       &	  4      &	  3          &  \citep{Scholz17ApJ}              \\	
    23	 & 2017/01/16   &  Eff/L-band	    &	16:14:20      &	14400	       &	  3	     &	  2	         &  \citep{Hardy17MN}                \\	
    24   & 2017/01/19   &  AO4.1-4.9GHz	  &	01:25:03      &	5901	       &	  1	     &	  0	         &  \citep{Michilli18Nat}	           \\	
    25	 & 2017/01/19   &  Eff/L-band	    &	16:05:40      &	12630	       &	  5	     &	  4	         &  \citep{Hardy17MN}                \\	
    26	 & 2017/01/25   &  Eff/L-band	    &	16:22:00      &	12600	       &	  4	     &	  3	         &  \citep{Hardy17MN}                \\	
	  27   & 2017/02/15	  &  AO/WE.G-1.38 GHz	&		          &		           &	  1	     &	  0          &  \citep{MAGIC18MN}                \\	
    28	 & 2017/02/19   &  Eff/L-band	    &	14:28:40      &	10800	       &	  1	     &	  0	         &  \citep{Hardy17MN}                \\	
	  19   & 2017/02/22	  &  AO/WE.G-1.38 GHz	&		          &		           &	  1	     &	  0	         &  \citep{MAGIC18MN}                \\	
	  30   & 2017/02/22	  &  AO/WE.G-1.38 GHz	&		          &		           &	  1	     &	  0	         &  \citep{MAGIC18MN}                \\	
	  31   & 2017/02/24	  &  AO/WE.G-1.38 GHz	&		          &		           &	  1	     &	  0	         &  \citep{MAGIC18MN}                \\	
	  32   & 2017/03/02	  &  AO/WE.G-1.38 GHz	&		          &		           &	  1	     &	  0	         &  \citep{MAGIC18MN}                \\
    33	 & 2017/08/26	  &  GBT/C-band     &	13:51:44	    &	1800	       &46$^{d}$   &   45          &  \citep{Gajjar18ApJ,ZhangYG18ApJ} \\	
    34	 & 2017/08/26	  &  GBT/C-band     &	14:21:59	    &	1800	       &47$^{d}$   &   46          &  \citep{Gajjar18ApJ,ZhangYG18ApJ} \\	
    35	 & 2017/08/31	  &  ARTS/L-wide	  &	06:23:37	    &	3600	       &	  1	     &	  0	         &  \citep{Oostrum17ATel}	           \\	 \hline
  \end{tabular}
  \begin{flushleft}
              Note. Each observation campaign is marked with a sequence number according to the start time, as shown in Column 1.
              Column 2 is the observation date; Column 3 presents the receiver and telescope name;
              Column 4 gives the start time of the observation;
              Column 5 gives observation duration;
              Column 6 gives the number of FRBs observed during the campaign, and Column 7 is the number of available waiting times;
              Column 8 provides the references.

               $^{a}$: In these observations, the source was monitored at AO/L-wide but the observations gave non-detection results.

               $^{b}$: In these observations, one burst was detected both by the VLA fast-dump observations at 3 GHz band and
                       Arecibo observations with L-wide receiver at a frequency range of 1.15 -- 1.73 GHz.

               $^{c}$: In these observations, the observation time was crossing, and two bursts were detected both by GBT/S-band
                       and AO/L-band receiver \citep{Scholz17ApJ}.

               $^{d}$: Observations by the GBT/4 -- 8 GHz(or C-band) receiver with Breakthrough Listen digital backend. Due to successful
                       application of deep learning \citep{ZhangYG18ApJ}, the number of bursts is tremendous
                       above 5.2 GHz.
                       Note that in these two observations, 21 bursts have been published earlier
                       by Gajjar et al. 2018 \citep{Gajjar18ApJ}. Here we only consider Zhang's results \citep{ZhangYG18ApJ}.
  \end{flushleft}
\end{table*}

\section{Statistical characteristics of waiting times }
We use the available waiting time data for statistical analysis. The distribution of the waiting time is plotted
as a histogram in Figure~1. It can be clearly seen that there is a bimodal distribution. A small portion of the
waiting times cluster in the range of 0.002 --- 0.04 s. A Gaussian fit for them gives
the peak time as $\Delta t \approx 0.0015 $ s. It is very striking that FRBs could repeat after
such a short quiescent period. However, note that due to very limited number of events,
significant fluctuation could be seen in the fit. In Figure~1, most of the waiting times lie
in the range of 1 --- 6100 s. They can be well fitted by another Gaussian function with a
reduced Chi-Sqr  $ \sim  6.259 $  and a Adj.R-Square of $ \sim  0.945 $.
The peak value of the waiting time is $\Delta t = 169.93$ for these events.
In our plot, since the X-axis is the logarithm of $\Delta t $, the Gaussian function means
these waiting times follow a Log-normal distribution.

\begin{figure}[htbp]
 \centering
  \includegraphics[width=0.5\textwidth]{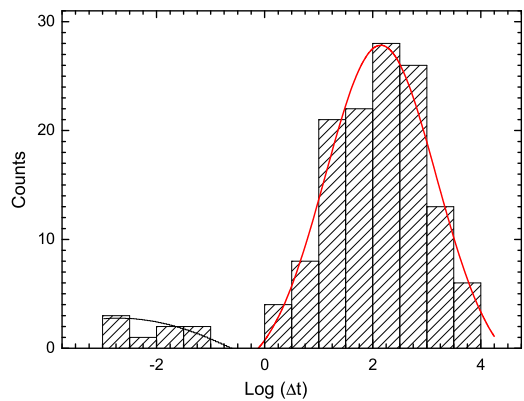}  \\
  \caption{A histogram plot of the distribution of the waiting time ($\Delta t$, in units of s).
           The X-axis of the $\Delta t$ data was logarithm, and the bin size is 0.5.
           A clear bimodal distribution can be seen. Most of the waiting times cluster at around several
           hundred seconds. They can be well fit by a Gaussian function, which means they follow a
           Log-normal distribution (the red line). A small portion of the waiting times are far below
           1 millisecond. They still could be fit with a Gaussian function (the black line), but
           with significant fluctuation.  }
  \label{fig1}
\end{figure}

The cumulative distribution of the waiting time is shown in the Figure~2. In this figure, the red line shows all the
calculated waiting time data. However, since the bursts were observed at different frequencies
by different receivers with different sensitivities, we have also plot the six sub-sets of data in the figure, i.e.
GBT/C-band (4-8 GHz), GBT/S-band (1.6--2.4 GHz), AO/C-band (4.1--4.9GHz), AO/L-band and ALFA (1.2--1.5 GHz) and
Eff/L-band (1.2--1.5 GHz), respectively. Generally speaking, these plots give us a straightforward vision of
the effect of the telescope sensitivity.
Due to the more efficient search technique of machine learning \citep{Gajjar18ApJ,ZhangYG18ApJ},
the contribution of the observation campaign at
GBT/C-band is very striking.

\begin{figure}[htbp]
 \centering
  \includegraphics[width=0.55\textwidth]{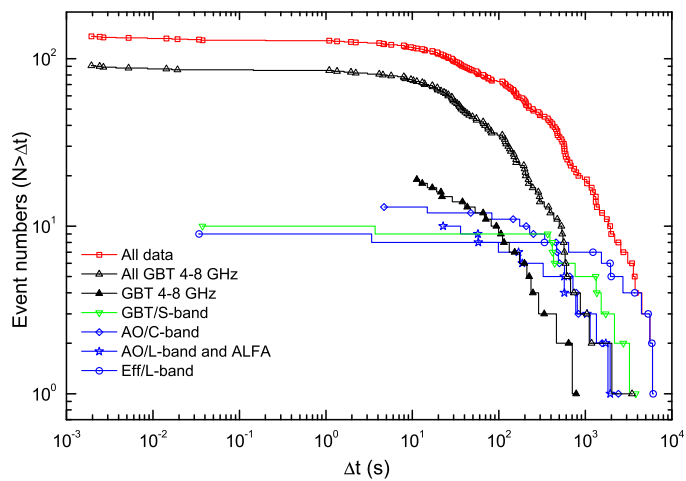}  \\
  \caption{Cumulative distribution of the waiting time ($\Delta t$).
           The hollow red squares show all the waiting time available for FRB 121102.
           Other plots correspond to various sub-set of the data, i.e.,
           the GBT/4-8 GHz data (hollow black triangles) \citep{ZhangYG18ApJ},
           part of the GBT/4-8 GHz data (filled black triangles) \citep{Gajjar18ApJ},
           the GBT/S-band data (opposite hollow triangles in green colour) \citep{Scholz16ApJ,Scholz17ApJ},
           the AO/C-band data (hollow diamonds in blue) \citep{Michilli18Nat},
           the AO/L-band and ALFA data (hollow stars in blue) \citep{Spitler16Nat,Scholz17ApJ},
           the Eff/L-band data (hollow circulars in blue) \citep{Hardy17MN}. }
  \label{fig2}
\end{figure}

According to some models, the energy released in an FRB burst are dependent on the quiescent period between
two consecutive bursts, i.e., the waiting time. This is especially true for FRB models in which some kinds
of intrinsic mechanisms are involved, such as the starquake models, the giant flare models of magnetars,
the accretion-induced collapse of the crust of neutron stars or strange stars, etc. The reason is obvious:
it is usually necessary for the system to accumulate some amount of energy to give birth to the next burst.
As a result, the intensity of the burst should be correlated with the waiting time.
On the contrary, in most models involving external mechanisms, no such correlation is expected. For example,
if the FRB is produced by the collision of a small body or by many small bodies in an asteroid
belt \citep{GengHuang2015ApJ,HuangYF16ASPC,Dai2016ApJ829}, the burst intensity will not correlate with
the waiting time. So, the relationship between the burst intensity and the waiting time can provide valuable
clues on the FRB models.

We have used our waiting time data to try to search for any possible correlation between the waiting time and
the burst intensity. However, before presenting the results, it should be further noted that the waiting
time actually could be correlated with either the preceding burst or the subsequent burst. We thus need to
check these two cases separately.

\begin{figure*}[htbp]
 \centering
  \includegraphics[width=1.0\textwidth]{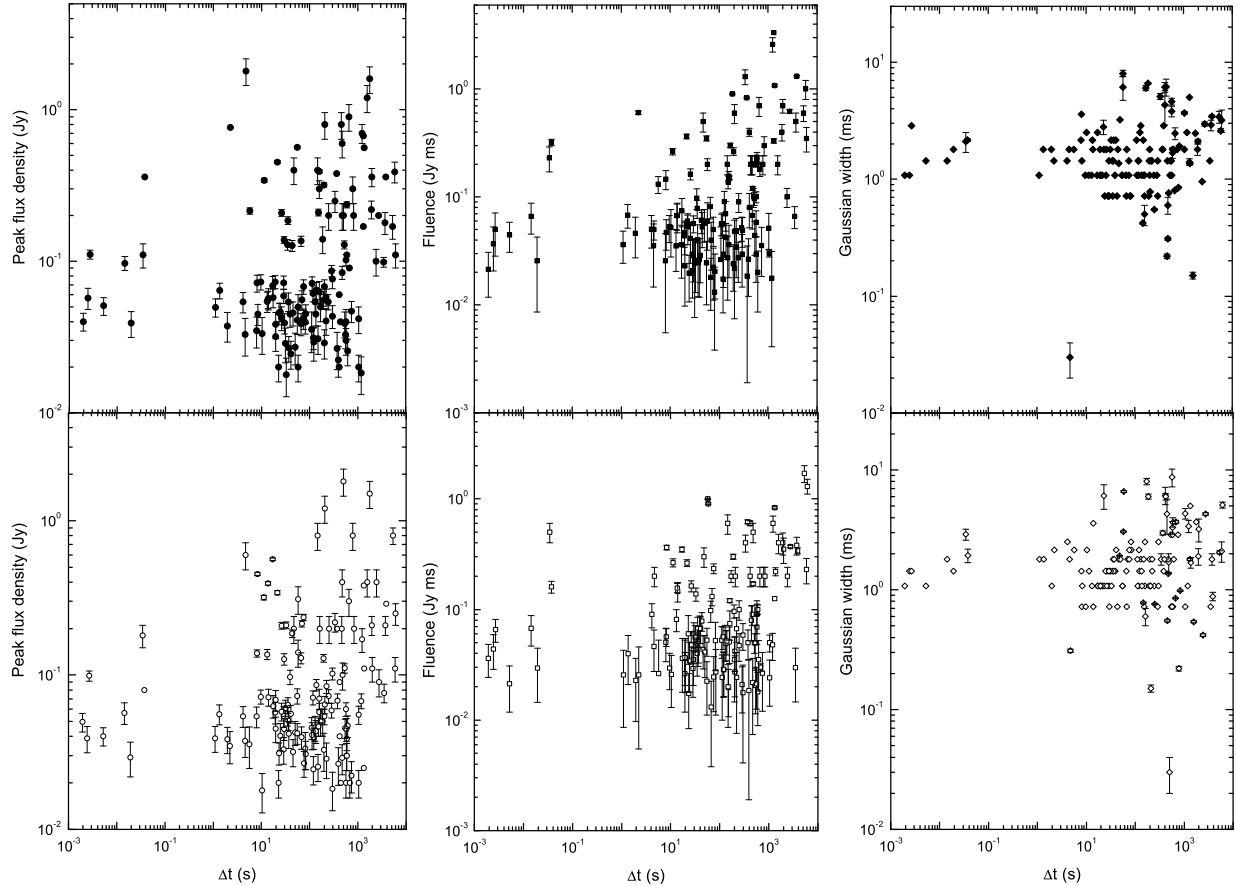}  \\
  \caption{Waiting time ($\Delta t$) versus other three parameters of FRB bursts, i.e., the
           peak flux density (the left panels), the fluence (the middle panels), and the width (the right panels).
           In the upper three panels, the parameters are for the preceding bursts, while in the bottom three panels, the parameters
           are for the subsequent bursts. In all the plots, no obvious correlation can be seen between
           the three parameters and the waiting time. }
  \label{fig3}
\end{figure*}

In Figure~3, we plot the waiting time versus some interesting parameters, such as the peak flux density ($S_{peak}$),
the duration (the Gaussian width, $w$), the fluence ($F$, which is calculated by multiplying $S_{peak}$ and $w$).
Form Figure~\ref{fig3}, we can see the data points are widely scattered in all the plots, so that
no obvious correlation can be observed between the burst intensity and the waiting time. This feature
can provide us important clues about the nature of of the repeating FRB source.

As mentioned in Section I, all the FRB models proposed in the literature can be grouped into
two different categories, i.e. catastrophic models and non-catastrophic models.
For FRB 121102, surely only the non-catastrophic models can meet its basic requirement of
repetition. It is interesting to note that the non-catastrophic models again can have two classes:
those due to internal mechanisms and those due to external mechanisms. For example, most of the
non-catastrophic models such as starquake-like explosions, flaring stars, lightning/wandering in pulsars, giant pulses/flares
from young pulsars, collapse of strange star crusts, are all mechanisms involving internal processes.
In these cases, the energy release is mainly caused by magnetic reconnection or accretion. So, to prepare for
a new burst event, it usually takes a period of time to accumulate the energy. As a result, there is
usually a positive correlation between the waiting time and the burst energetics.
On the contrary, models involving the collisions between small bodies and neutron
stars \citep{GengHuang2015ApJ,HuangYF16ASPC,Dai2016ApJ829}
are the few kinds of non-catastrophic models in which external mechanisms are engaged.
In these cases, the happening of the bursts is almost completely random. The system does
not need to accumulate any intension for the next burst.
Therefore, the nonexistence of any correlation between the waiting time and other parameters (especially
with the burst energetics) as shown in our Figure~\ref{fig3} seem to rule out many intrinsic
mechanism models for FRBs, but strongly support the external mechanism models, especially the collisions between small bodies and neutron
stars \citep{GengHuang2015ApJ,HuangYF16ASPC,Dai2016ApJ829}.

\section{Discussion and Conclusion}

FRB 121102 is the only source that is observed to burst out repeatedly till now. In this study, we analyze
the repeating behavior of FRB 121102 statistically, paying special attention on the waiting time.
It is found that the waiting time shows a clear bimodal distribution, clustering at 0.002 --- 0.04 s and
170 s respectively. It is striking to note that some bursts could happen very shortly after the preceding
burst. We also tried to examine any possible correlation between the waiting time and the burst intensity.
It is found that the waiting time does not correlate with the intensity (characterized by the peak flux and
the burst fluence) of either the preceding burst or the subsequent burst. This result strongly indicates that
the repeating bursts should be produced by some external mechanisms, but not intrinsic mechanisms.
We suggest that the models involving collisions between small bodies and
neutron stars \citep{GengHuang2015ApJ,HuangYF16ASPC,Dai2016ApJ829} could be competitive mechanisms for FRBs.

Currently, due to the occultation of the Earth, FRB 121102 could not be continuously monitored for too long in any
observational campaign. Usually, it could be continuously monitored for only one or two hours, and in a few very
rare cases, the monitoring period could be extended to several hours. As a result, it is difficult for us to
derive the waiting time that is longer than several thousands. This may be the reason that there is a cutoff
at the longer section of the waiting time distribution. It is suggested the various large telescopes could
cooperate to carry out an all-time monitoring campaign on FRB 121102. It can help to clarify the waiting
time distribution in the longer period regime, and may be important for understanding the nature of this
enigmatic FRB source.

The bursts of FRB 121102 shows a clustering behavior in time. There are many outbursts in some particular
time peroid, which leads to very short waiting time for these bursts.
For example, six bursts (Bursts 6 to 11) arrived in less than twenty minute period and four bursts
(Bursts 13 to 16) occurred in a $\sim 50 $ minute period during the Observation Campaign 4.
Even more prominent is on the 33th and 34th observation campaign, where more frequent eruptions
were observed within half an hour. It indicates that there appears to be epochs in which the
source is more active. However, it should also be pointed out that there are also several very
long observing sessions which resulted in non-detections \citep{Scholz16ApJ,Law17ApJ,Price18RN}.
The clustering phenomenon has been discussed by a few authors \citep{Oppermann2017arX}.
For example, a neutron star traveling through an asteroid belt can naturally lead to an active period and
give birth to such a phenomenon \citep{GengHuang2015ApJ,HuangYF16ASPC,Dai2016ApJ829}.
In this case, the next active period could even be forcasted \citep{Bagchi17ApJ}.
However, since FRB 121102 has been discovered for only less than 8 years, it is still an open question
whether the repeating activity shows any difference on larger time-scales.

As shown in our Figure~1, the majority of the waiting times follow a log-Gaussian distribution.
This behavior is very similar to a few activities observed in other celestial objects, such as
the waiting times of hard X-ray bursts from the Sun \citep{Wheatland1998ApJ}, the waiting
times of hard X-ray bursts from the famous soft gamma-ray repeater SGR $1900+14$ \citep{G1999ApJ}, and
the waiting times of the Crab pulsar glitches ($\sim 10^{2}$ to $\sim 10^{6}$ s )\citep{Haskell2016MN}.
In all these cases, the waiting times follow similar log-normal distributions.
The common reason may be that they are all random processes to some extent.

In the bimodal distribution of the waiting time, there are about ten bursts clustered in the region with
the waiting time less than 1 second, most of which were found by using the new
searching technique \citep{ZhangYG18ApJ}.
However, while a few of these bursts are weak, there are also some strong bursts among them.
It is thus even more difficult for the intrinsic mechanisms to explain such a fact: how the
central engine could produce so many strong bursts in such a short period. On the contrary,
this issue can be naturally accounted for by an external mechanism such as the collision
between small bodies with neutron stars. It is possible that when a neutron star travels
through an asteroid belt, it may encounter several asteroids one after the other, producing
a group of bursts in a short time.

\section*{Acknowledgments}

This work is supported by the National Natural Science Foundation of China (Grant Nos. 11873030, 11473012, and U1431126),
the National Postdoctoral Program for Innovative Talents (Grant No. BX201700115),
China Postdoctoral Science Foundation funded project (Grant No. 2017M620199),
Guizhou Provincial Natural Science Foundation (201519),
and by the Strategic Priority Research Program of the Chinese Academy of Sciences
``Multi-waveband Gravitational Wave Universe'' (Grant No. XDB23040000 and XDB23040400).
LMS acknowledges support from the National Program on Key Research and Development Project
(Grant No. 2016YFA0400801). YPY is supported by the Initiative Postdocs Supporting
Program (No. BX201600003) and the China Postdoctoral Science Foundation (No. 2016M600851).
BL is also supported by Key Research Program of Frontier Sciences, Chinese Academy of
Sciences (Grant NO. QYZDB-SSW-SLH012).


\end{document}